\documentclass[sigconf,nonacm]{acmart}
\usepackage{hyperref}
\usepackage{algorithmic}
\usepackage{textcomp}
\usepackage{todonotes}
\usepackage{url}
\usepackage{enumitem}
\usepackage{xspace}
\usepackage{soul}
\usepackage{comment}
\usepackage{stfloats}
\usepackage{tabularx}
\usepackage{booktabs}
\usepackage{caption}
\usepackage{xurl}
\usepackage{listings}
\usepackage{diagbox}
\usepackage{pifont}
\usepackage{array}
\usepackage{colortbl}
\usepackage{threeparttable}

\newcommand{\cmark}{\ding{51}} 
\newcommand{\xmark}{\ding{55}} 

\definecolor{lightgrayrow}{gray}{0.92}

\newcommand{\DANY}[1]{\sethlcolor{green}\hl{\textbf{Dany: } #1}}

\newcommand{\ctg}{\textsc{CTG}\xspace}
\newcommand{\botium}{\textsc{Botium}\xspace}
\newcommand{\charm}{\textsc{Charm}\xspace}
\newcommand{\asymobtcg}{\textsc{Asymob TCG}\xspace}
\newcommand{\tracer}{\textsc{Tracer}\xspace}

\newcommand{\datasetR}{\textsc{BRASATO}\xspace}

\newcommand{\datasetD}{\textsc{COD}\xspace}

\newcommand{\newtext}[1]{\textcolor{black}{#1}}

\setcopyright{acmlicensed}
\copyrightyear{2026}
\acmYear{2026}
\acmDOI{XXXXXXX.XXXXXXX}
\acmConference[MSR '26]{23rd International Conference on Mining Software Repositories}{April 13--14, 2026}{Rio de Janeiro, Brazil}
\acmISBN{978-1-4503-XXXX-X/2018/06}

\begin{document}

\title{Automated Testing of Task-based Chatbots: How Far Are We?}

\author{Diego Clerissi}
\email{diego.clerissi@unimib.it}
\orcid{0000-0001-7651-0400}
\affiliation{
  \institution{University of Milano-Bicocca}
  \city{Milan}
  \country{Italy}
}

\author{Elena Masserini}
\email{elena.masserini@unimib.it}
\orcid{0009-0002-6969-1500}
\affiliation{
  \institution{University of Milano-Bicocca}
  \city{Milan}
  \country{Italy}
}

\author{Daniela Micucci}
\email{daniela.micucci@unimib.it}
\orcid{0000-0003-1261-2234}
\affiliation{
  \institution{University of Milano-Bicocca}
  \city{Milan}
  \country{Italy}
}

\author{Leonardo Mariani}
\email{leonardo.mariani@unimib.it}
\orcid{0000-0001-9527-7042}
\affiliation{
  \institution{University of Milano-Bicocca}
  \city{Milan}
  \country{Italy}
}

\renewcommand{\shortauthors}{Clerissi et al.}

\begin{abstract}
Task-based chatbots are software, typically embedded in real-world applications, that assist users in completing tasks through a conversational interface. 
As chatbots are gaining popularity, effectively assessing their quality has become crucial. 
Whereas traditional testing techniques fail to systematically exercise the conversational space of chatbots, several approaches specifically targeting chatbots have emerged from both industry and research. Although these techniques have shown advancements over the years, they still exhibit limitations, such as simplicity of the generated test scenarios and weakness in implemented oracles. 
In this paper, we conduct a confirmatory study to investigate such limitations by evaluating the effectiveness of state-of-the-art chatbot testing techniques on a curated selection of  task-based chatbots from GitHub, developed using the most popular commercial and open-source platforms. 
\end{abstract}

\maketitle

\section{Introduction}\label{sec:introduction}
Conversational systems are increasingly being adopted in a wide range of domains~\cite{adamopoulou2020chatbots}, from customer support to digital public services, e-commerce, and information delivery platforms~\cite{cui2017superagent,laranjo2018conversational,fiore2019forgot,ukpabi2019chatbot,de2021s}. Among them, task-based chatbots represent a prominent class designed to assist users in completing specific tasks through structured dialog flows~\cite{kucherbaev2018human,grudin2019chatbots,adamopoulou2020chatbots}, as opposed to LLM-based chatbots that, although receiving significant attention, still lack the degree of control necessary to deliver reliable services to end-users in several domains. As these systems are deployed in the real-world and interact with end-users and external services, ensuring their correctness and reliability becomes essential.

Testing conversational systems, however, introduces specific challenges compared to traditional software~\cite{cabot2021testing,li2022review,lambiase2024motivations}. These systems combine a conventional software layer responsible for business logic, data handling, and service integration with a conversational interface that interprets and manages natural-language interactions. This dual nature makes them particularly difficult to test systematically. For instance, test case generation techniques must generate conversations that are thorough enough to activate the inner procedures and services implemented by the chatbots, and must also be able to interpret the natural-language responses to establish the outcome of the tests.

With a growing research and industrial interest in conversational software, test case generation tools for task-based chatbots are appearing. Existing approaches primarily focus on automating conversational scripts and simulating faults. For example, \botium~\cite{botium} can generate tests that sample the main conversational flows of a chatbot, while \charm~\cite{bravo2020testing} and \ctg (Chatbot Test Generator)~\cite{rapisarda2025test} augment \botium with additional test generation capabilities. Mutation testing has also been considered to assess test case generation techniques \cite{ferdinando2024mutabot,gomez2024mutation,clerissi2025towards}.

However, these approaches are limited in their generation and verification capabilities. In fact, they are unable to explore multi-step conversations, being mostly limited to single request-response interactions. Moreover, the oracle is essentially limited to the regression oracle, with strong difficulties in handling any flexibility in the responses. 

Motivated by these observations derived from our preliminary empirical experiments on chatbot testing~\cite{ferdinando2024mutabot,rapisarda2025test,masserini2025brasato,clerissi2025towards,masserini2026assessing}, we formulate the following hypothesis:
\textit{Current chatbot testing techniques are unable to comprehensively assess the behavior of task-based chatbots and lack reliability in generating correct and robust tests.}

To investigate these limitations thoroughly and systematically, we propose to conduct a \emph{confirmatory study} based on (1) the definition of a \emph{large curated set of task-based chatbots}, developed with different platforms and technologies, and (2) the \emph{experimentation of state-of-the-art test generation approaches} with the selected chatbots.


In  particular, our investigation aims to target the following key research questions: 

\textbf{RQ1 (Test Correctness):
To what extent can existing automated testing tools generate executable test cases?} Generating tests as conversations is a new challenge, and sometimes techniques struggle to generate fully correct test cases. This research question investigates this aspect, studying the factors that might lead to the generation of incorrect test cases that cannot be executed \newtext{or that, although executable, cannot be interpreted by the chatbot}.

\textbf{RQ2 (Conversations):
To what extent can existing automated testing tools effectively explore the conversational space of task-based chatbots?} This RQ investigates the capability of test case generation tools to sample the conversational space and cover the conversational flows supported by the chatbot under test. Missing relevant conversational flows implies missing to cover both the natural-language pipeline implemented to handle the missed flows and the backend features possibly activated by those flows. It is thus important to assess the capabilities of automated tools to identify the most relevant directions of improvement for the future.

\textbf{RQ3 (Functions): To what extent do automated conversational tests exercise the underlying functionality of task-based chatbots?} This RQ investigates the capability of test case generation tools to fully exercise a chatbot, ultimately and comprehensively activating the backend features through conversations. The inability of tools to reach backend features may result in important bugs remaining unnoticed.

\textbf{RQ4 (Oracle): How reliable are automated conversational oracles in detecting failures?} Sampling conversations and backend features is effective only as long as tests can accurately reveal failures. This also depends on the strength of the oracles used to judge executions~\cite{barr2014oracle}. When dealing with conversations, the oracle problem is exacerbated by the ambiguity of the language; that is, not only the responses can be hard to predict, but they can also be stated in a multitude of ways. Thus, interpreting the responses produced during a conversation is a hard problem for test case generation techniques. This RQ investigates how strong the oracles generated by state-of-the-art approaches are.

\textbf{RQ5 (Flakiness): How flaky are conversational tests?} Tests must ideally produce the same result every time they are executed. However, tests might be flaky, making test execution outcomes brittle.
This RQ empirically investigates the flakiness of conversational tests, measuring the magnitude and causes of this phenomenon.

RQ1-5 collectively provide, on one hand, a comprehensive view on the effectiveness of testing tools, and on the other hand, insights on the reliability of open-source task-based chatbots.


\newtext{To address this investigation, we start from the \datasetR~\cite{masserini2025brasato} and the \datasetD ~\cite{masserini2026assessing} datasets. \datasetR is a curated dataset of 193 Rasa~\cite{rasa} task-based chatbots available on GitHub. \datasetD similarly includes 185 curated Dialogflow~\cite{dialogflow} task-based chatbots. Among the platforms for the development of task-based chatbots, Rasa and Dialogflow are the most popular open-source and commercial solutions~\cite{abdellatif2021comparison,motger2022software,cao2023characterizing,perez2021choosing,abdellatif2021comparison,benaddi2024systematic}, respectively. 
To mitigate potential platform-specific biases (e.g., Rasa chatbots are all implemented in Python, while Dialogflow chatbots are tightly coupled with proprietary Google services) and improve the generalizability of our findings, as a first step, we plan to extend our curated selection of task-based chatbots to Amazon Lex~\cite{amazon-lex}, that is another popular commercial solution~\cite{perez2021choosing,benaddi2024systematic}, replicating the same methodology used to build \datasetR and \datasetD. In this way, we will extend our empirical investigation to chatbots implemented 
considering a multitude of languages 
and implementation options.}

We identified \emph{five} state-of-the-art test case generation techniques for task-based chatbots that we plan to use to answer our research questions: \botium~\cite{botium}, which is a multi-platform state-of-the-practice test generation tool for task-based chatbots; \charm~\cite{bravo2020testing}, which extends \botium with the capability to generate robustness tests; \ctg~\cite{rapisarda2025test}, which is a dynamic test generation tool that analyzes chatbot responses during test generation; \asymobtcg (Test Case Generator)~\cite{canizares2024coverage}, which targets dialog coverage metrics as a goal of the test generation process; and finally \tracer~\cite{del2025automated}, which employs LLMs to synthesize test profiles from interactions with the chatbot and uses them for user simulation.

In the rest of this paper, we describe the core elements of a task-based chatbot (Section ~\ref{sec:background}) and the structure of conversational tests (Section~\ref{sec:conv}), we present the methodology that we designed to answer the five RQs (Section~\ref{sec:study}), we discuss the threats to validity (Section~\ref{sec:threats}) and the related work (Section~\ref{sec:related}), and we finally provide some remarks (Section~\ref{sec:conclusions}). 

\section{Task-Based Chatbots}\label{sec:background}
Task-based chatbots are software systems designed to guide users through structured interactions to accomplish specific goals, such as booking travel, ordering food, or handling customer service requests~\cite{kucherbaev2018human,grudin2019chatbots,adamopoulou2020chatbots}. These systems follow predefined conversational flows and often integrate with backend services to complete tasks.

\begin{figure*}[t!]
    \centering
\includegraphics[width=0.7\linewidth, trim=0.2cm 8.4cm 0.5cm 0.0cm, clip=true]{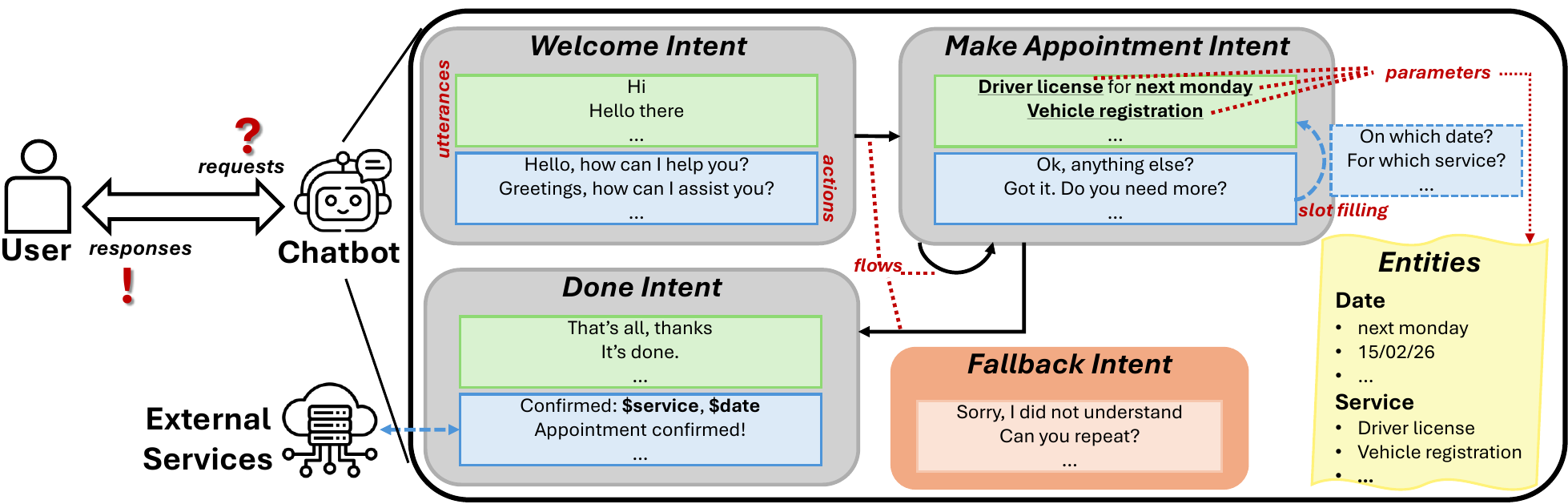}
    \caption{Task-based chatbot architecture.}
    \label{fig:chatbot}
\end{figure*}

Figure \ref{fig:chatbot} exemplifies the internal architecture of a generic task-based chatbot.
Task-based chatbots implement one or more \textit{intents}, which represent the goals that users may want to achieve by interacting with chatbots (e.g., the \texttt{Make Appointment} intent represents the goal of defining the desired service and data of an appointment). Each intent includes a set of \textit{utterances} and \textit{actions}. Utterances represent the phrases used to train the chatbot in recognizing the associated user intent (e.g., ``Driver license for next monday'' is an example utterance for the \texttt{Make Appointment} intent). \textit{Actions} can be either plain-text responses (e.g., an action that returns ``Ok, anything else?'' to a user request for an appointment) or advanced operations that execute custom implementations \newtext{of the business logic and interactions with} external services (e.g., the \texttt{Done} intent including an action that returns ``Appointment confirmed!'' after invoking the Google Calendar API to create an event). 

Chatbots also define \textit{entities}, which are the datatypes a chatbot can understand (e.g., \texttt{Date} values). Entities can be instantiated in a conversation by means of \textit{parameters} within the utterances (e.g., the values `Driver license' for the \texttt{Service} entity and `next monday' for the \texttt{Date} entity in the first utterance of the \texttt{Make Appointment} intent) and referred in the chatbot actions (e.g., \texttt{\$service} and \texttt{\$date} in the first action of the \texttt{Done} intent). 
When required parameters are missing in a user request, the \textit{slot filling} mechanics is activated, if implemented, to ask specific questions to the user to elicit them (e.g., asking ``On which date?'' if the user requests an appointment without proposing a date).
Intents are connected to compose the \textit{flows}, which represent the possible conversational paths, by interleaving user requests and chatbot responses. 
A \textit{fallback} intent can be defined to handle the cases in which the chatbot is unable to understand a user request, asking the user to rephrase it.

An interaction with a chatbot starts with a user submitting a request. The chatbot uses NLP (Natural Language Processing)  models trained with the utterances to associate the request with an intent. Based on the detected intent, the chatbot produces a response. If the intent is associated with any action, the action is executed, passing the parameters extracted from the request, if needed. If no intent is recognized, the fallback intent is executed. This interaction continues until the user is satisfied with the response. 

\section{Conversational Tests}\label{sec:conv}
Conversational tests are tests that exercise the functionalities of a task-based chatbot through conversations. Similarly to other types of tests, conversational tests consist of a sequence of inputs (i.e., the natural language requests issued by the user) and expectations (i.e., a specification of the expected natural language response of the chatbot).

\begin{figure}[t!]
\centering
\includegraphics[width=0.9\columnwidth, trim=0.0cm 0.0cm 16.5cm 0.0cm, clip=true]{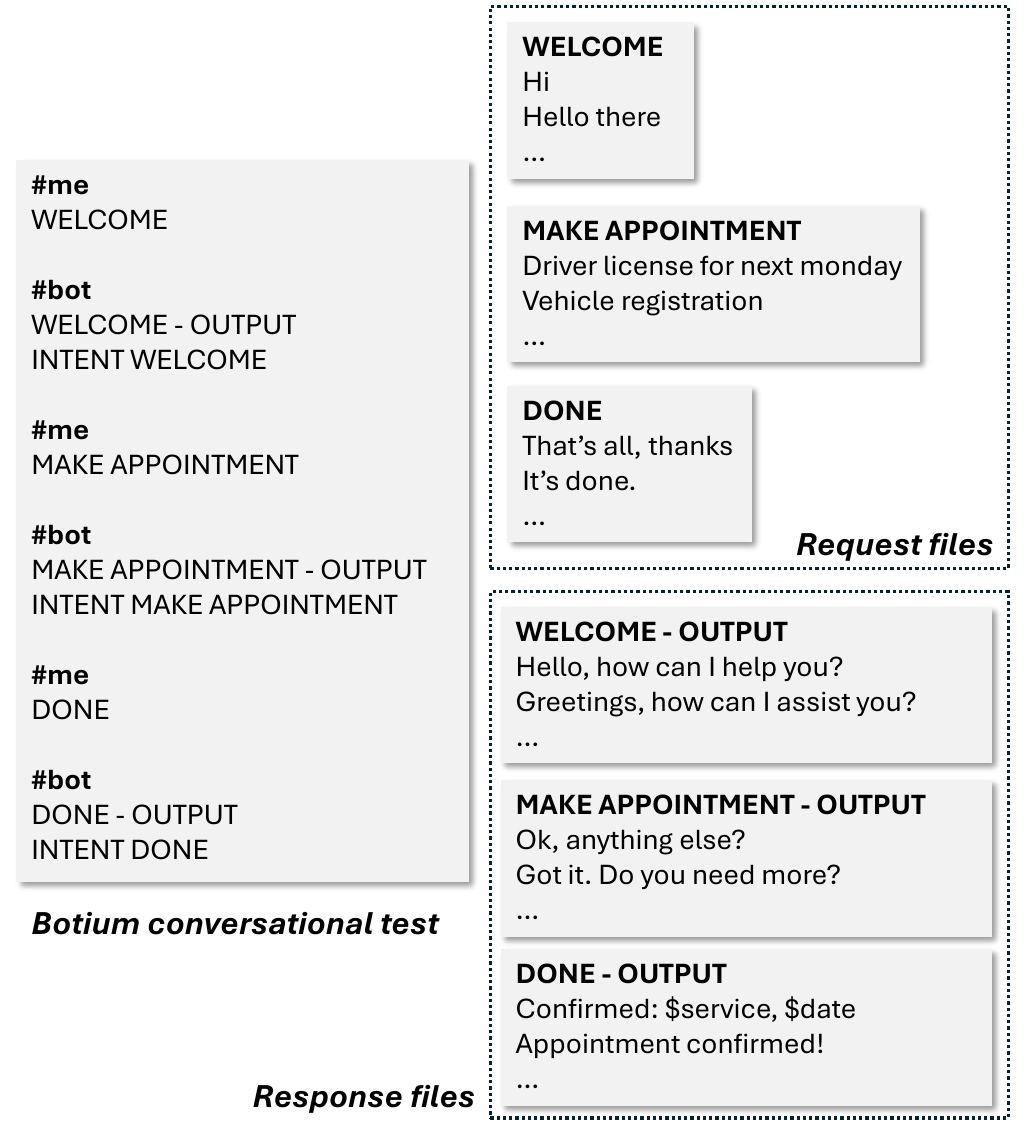}
\caption{A \botium conversational test.}
\label{fig:test}
\end{figure}

Figure~\ref{fig:test} shows how a conversational test for the example chatbot, depicted in Figure~\ref{fig:chatbot}, is defined in \botium. \botium is a multi-platform quality assurance framework that can automatically generate and execute conversational tests defined as a sequence of \#me and \#bot interactions. The \#me sections represent the requests a user can send to a chatbot, whereas the \#bot sections represent the expected responses. 


Since the same intent can be activated by multiple user requests formulated in different ways, and the same intent may generate different chatbot responses for the same input request, \botium uses external files as references for the \#me and \#bot sections. These files list the many ways the same requests and responses can be stated. 
For instance, the \texttt{WELCOME} label following the first \#me interaction in Figure~\ref{fig:test} refers to the homonymous request file, which includes utterances such as ``Hi'', ``Hello there'', and more, whereas the \texttt{WELCOME - OUTPUT} label following the first \#bot interaction refers to the homonymous response file, which lists the expected textual responses, such as the string ``Hello, how can I help you?''.

During test execution, the references to the request files are expanded and separately sent to the chatbot. \botium tries every possible combination of values, but other tools, such as \ctg, implement alternative strategies to avoid the combinatorial explosion of the tests. Once the chatbot responds, the response is matched against those included in the response file specified in the \#bot section. The set of responses in the file works as a flexible test oracle: the response is correct if it matches at least one of the entries in the file; otherwise, the test fails. 

Additionally, conversational tests may also implement oracles not based on the actual response that the chatbot returns. This is useful when the actual response of the chatbot cannot be predicted. \botium supports multiple asserters that can be used to this end\footnote{\url{https://botium-docs.readthedocs.io/en/latest/05_botiumscript/index.html\#using-asserters}}. A notable option is the specification of the intent that a request is expected to activate. This oracle can be simply implemented using the keyword \texttt{INTENT} followed by the name of the intent that must be activated. If a different intent is activated, the test fails. 

In the example test in Figure~\ref{fig:test}, every interaction includes both the expected textual response and the intent oracle. However, a conversational test may specify just one of them or even none of them. In the latter case, any response produced by the chatbot and any intent that is activated would be fine from the perspective of the test. 

\begin{figure}[t!]
\centering
\includegraphics[width=\linewidth, trim=0.0cm 5.7cm 15.0cm 0.0cm, clip=true]{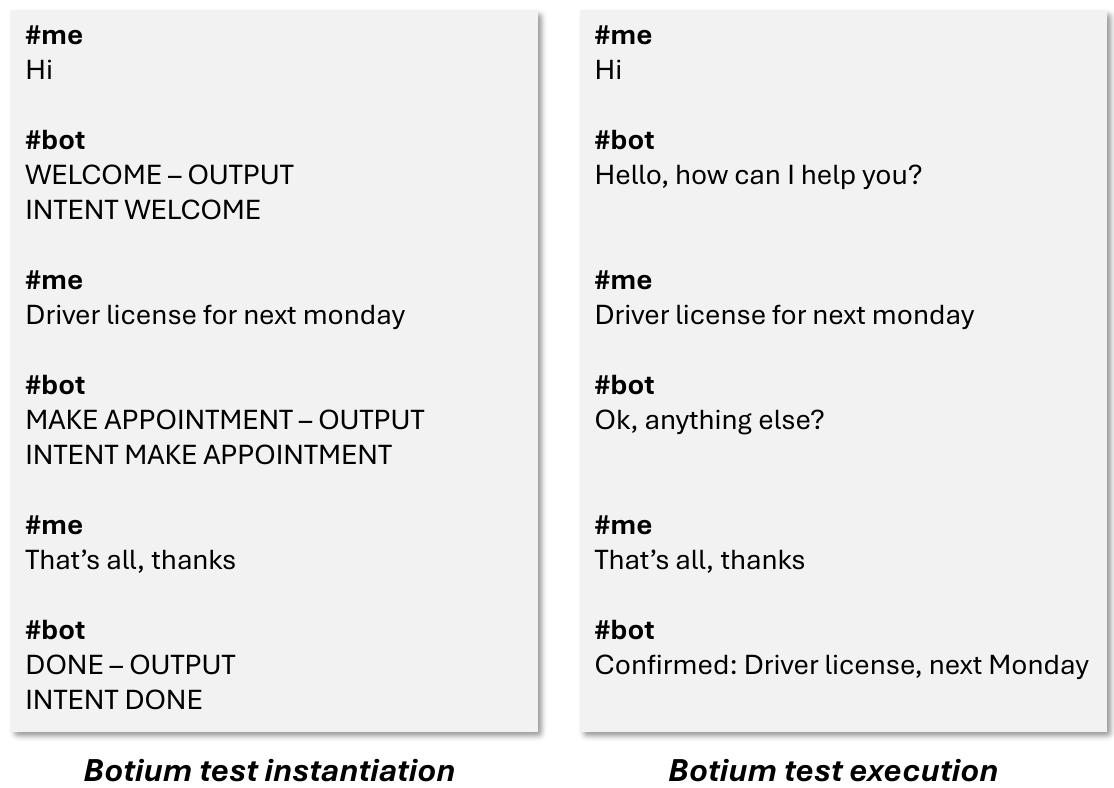}
\caption{A \botium test instantiation and execution.}
\label{fig:convo}
\end{figure}

Figure~\ref{fig:convo} shows an instantiation of the conversational test presented in Figure~\ref{fig:test} (left-hand side) and an example of actual conversation that can be observed when the test is executed (right-hand side).

The instantiation is the same as the test in Figure~\ref{fig:test}, with the only difference that the user interactions are explicitly defined, that is, actual sentences are selected from the request files and used to replace the filename in the test. The execution is the log of the submitted requests and the returned responses.  

In the first user-bot interaction, the user and the chatbot greet each other (\texttt{WELCOME} intent). Then, in the second user-bot interaction, the user specifies the service (``Driver license'') and the date (``next monday''), providing values to the required parameters (i.e., the \texttt{service} and the \texttt{date} parameters, see Figure~\ref{fig:chatbot}). The chatbot saves the parameter values internally and asks if anything else is needed (\texttt{MAKE APPOINTMENT} intent). Finally, in the last user-bot interaction, the user confirms that nothing more is needed, and the chatbot confirms the appointment by listing the values of the formerly provided parameters (\texttt{DONE} intent). The confirmation also activates the execution of an action that involves the Google Calendar service to book the appointment. The execution of the action is only indirectly observable by the user, since it is implicit in the confirmation of the appointment.  

The responses of the chatbots are assessed in accordance with the \#bot interactions specified in the instantiated test. In this example case, all the responses are consistent with the expectation, both as possible sentences and activated intent, and thus the test passes.


\section{Design of the Experiments}\label{sec:study}
In this section, we describe how we define the dataset that will be used in the study, we discuss how we selected the test generation tools that will be experimented on the selected chatbots, and we detail the experimental procedure followed to answer the research questions presented in Section~\ref{sec:introduction}.

\subsection{Dataset}

We plan to conduct our investigation considering relevant and representative open-source task-based chatbots. We focused on the most popular task-based platforms for the study. In particular, Rasa, Google Dialogflow, Amazon Lex, Microsoft Bot, and IBM Watson are systematically acknowledged as the most popular platforms~\cite{perez2021choosing,abdellatif2021comparison,benaddi2024systematic}. Since Microsoft Bot has been recently discontinued and IBM Watsonx free plan is insufficient for  experimentation, we targeted the remaining three platforms.


\newtext{Regarding Rasa and Dialogflow, we start from the \datasetR~\cite{masserini2025brasato} and ~\datasetD\cite{masserini2026assessing} datasets. \datasetR is a curated dataset of 193 Rasa chatbots, while \datasetD is a curated dataset of 185 Dialogflow chatbots, all selected according to criteria about the complexity of the conversations they can handle, the implemented functionalities, the integration with external services, and the integration with the most recent version of the platforms.}


\newtext{We plan to replicate the same methodology used to build \datasetR and \datasetD to cover also the chatbots developed with Amazon Lex.} 
In particular, we search for GitHub repositories containing the name of the target platform (e.g., "Rasa") and the keyword "chatbot" in the metadata files (e.g., in the README), using the GitHub Repository API\footnote{https://docs.github.com/en/rest/repos}. The set of retrieved repositories is analyzed, retaining only those containing syntactically correct chatbots (e.g., those that can be instantiated on their platform).  
We analyze the identified chatbots to extract conversational parameters (e.g., list of intents and entities) and supported languages, removing any duplicates based on parameter equivalence and code similarity (with 95\% threshold, to accommodate for trivial differences, such as code comments). 
The curated dataset is obtained by filtering for dialog complexity, functional complexity, and utility, retaining chatbots with at least one intent, one entity, one custom backend action, support for the English language, developed for the most recent version of the platform, and having stars assigned on GitHub. 


\newtext{Compared to Rasa and Dialogflow chatbots, Amazon Lex chatbots have distinguishing characteristics that may require some minor adaptations of the methodology. For instance, the range of languages that can be used to implement chatbots and the architecture of the chatbots, although including the same key elements as discussed in Section~\ref{sec:background}, is different.}

The overall set of chatbots collectively included in the three datasets is expected to be prohibitively large, 
considering that experiments will require deploying and executing every subject chatbot. 
We plan to work with a target set of $45$ chatbots, $15$ chatbots per platform, to ensure sufficient diversity and representativeness. Although ambitious, we estimate that working with so many chatbots is feasible, based on our previous experience~\cite{masserini2025brasato}. Moreover, a study with 45 chatbots that is three times the largest study appeared so far on testing of task-based chatbots~\cite{gomez2024mutation}, combined with the largest set of test generation tools used (see Section~\ref{sec:testgen}), results in a study of highly relevant magnitude. 


We will select the subjects from each curated dataset by applying the following iterative procedure. First, we classify chatbots according to their domain (e.g., medical, education, finance). This information is readily available for the chatbots in \datasetR \newtext{and \datasetD}. The classification has been derived by asking ChatGPT-4o to map each chatbot to a Google Play Category\footnote{\url{https://www.searchapi.io/docs/parameters/google-play-store/categories}} based on the chatbot conversational parameters and the repository title, description, and README file. We will repeat the same process for the chatbots in the 
\newtext{Amazon Lex dataset}.
Then, we will perform a stratified selection across domains of the chatbots that provide the highest number of intents, process the highest number of parameters, and 
\newtext{implement the highest number of actions that include concrete business logic and interactions with external services.}
This process accounts for conversational and functional complexity, while guaranteeing domain diversity. 
If, at any stage of the experimentation, a chatbot is found broken (e.g., because a third-party service used is no longer available), we will attempt to fix it (e.g., by replacing the obsolete service with an equivalent one), as long as the fixing does not require re-implementing features. If fixing is not possible, we will replace the chatbot with another one from the curated dataset that is aligned with the aforementioned complexity and diversity criteria. 
If no more candidates are available, to reach the intended subject size, we will resort to the larger datasets of all chatbots per platform available on GitHub, from which the curated datasets are derived~\cite{masserini2025brasato,masserini2026assessing}. As the curated datasets are derived using, among the criteria, a threshold on the number of stars, we will drop this criterion, including less popular chatbots that still satisfy the rest of the criteria, in the experimentation. Considering the number of chatbots likely available in the curated datasets (e.g., 193 for Rasa), this criterion is unlikely to be necessary.

\subsection{Test Generation Tools} \label{sec:testgen}


We selected the state-of-the-art test generation tools by systematically analyzing the literature. In particular, we submitted the query \textit{``chatbot testing'' AND (``task-based'' OR ``task-oriented'')} to Google Scholar in January 2026, to extract papers that address the testing of task-based chatbots. 
We obtained a total of 73 papers that we manually inspected to eliminate the irrelevant ones, 
according to the following exclusion criteria: (i) not being written in English; (ii) not having chatbot testing as the main topic; (iii) not being a peer-reviewed publication (e.g., master's thesis, arXiv). We ended up with 7 papers, then we performed iterative snowballing to collect any additional paper. 
We finally obtained a total of 19 papers, which we more carefully inspected to collect any presented or used task-based chatbot testing tool. We excluded from our experiment the tools that match any of the following exclusion criteria: (i) not being an End-to-End testing tool for task-based chatbots; (ii) not providing free access to the tool; (iii) not supporting at least one platform among Rasa, Google Dialogflow, and Amazon Lex. 
We finally selected five tools, those satisfying all criteria, as shown in Table \ref{tab:all-tools}:

\begin{table}[t]
\centering
\footnotesize
\setlength{\tabcolsep}{4pt}
\renewcommand{\arraystretch}{1.15}
\caption{Chatbot-related tools.}
\begin{tabular}{p{3.1cm}|c|c|c}
\hline
\textbf{Tool} & \textbf{E2E Testing} & \textbf{Accessible} & \textbf{Platforms} \\
\hline
Alexa Test Skill Framework~\cite{alexa-test-framework} & \xmark & \cmark & \xmark \\
\rowcolor{lightgrayrow}
\asymobtcg~\cite{canizares2024coverage} & \cmark & \cmark & \cmark \\
Bespoken~\cite{bespoken} & \cmark & \xmark & \cmark \\
Botanalytics~\cite{botanalytics} & \xmark & \cmark & \cmark \\
BoTest~\cite{ruane2018botest} & \cmark & \xmark & \xmark \\
\rowcolor{lightgrayrow}
\botium~\cite{botium} & \cmark & \cmark & \cmark \\
Bot Tester~\cite{vasconcelos2017bottester} & \cmark & \cmark & \xmark \\
Bo{\v{z}}i{\'c} Metamorphic~\cite{bovzic2022ontology} & \cmark & \xmark & \cmark \\
Bo{\v{z}}i{\'c} \textit{et al.} AI Planning~\cite{bozic2019chatbot} & \cmark & \xmark & \cmark \\
\rowcolor{lightgrayrow}
\charm~\cite{bravo2020testing} & \cmark & \cmark & \cmark \\
Chatbottest~\cite{chatbottest} & \xmark & \cmark & \xmark \\
\rowcolor{lightgrayrow}
\ctg~\cite{rapisarda2025test} & \cmark & \cmark & \cmark \\
DialTest~\cite{liu2021dialtest} & \xmark & \cmark & \xmark \\
LemmaCartridge~\cite{yalla2020ai} & \cmark & \xmark & \xmark \\
\rowcolor{lightgrayrow}
\tracer~\cite{del2025automated}  & \cmark & \cmark & \cmark \\
\hline
\end{tabular}
 \begin{tablenotes}
 \footnotesize
 \item[a)] \textit{E2E Testing}: the tool generates End-to-End conversational tests. \textit{Accessible}: the tool is accessible for use. \textit{Platforms}: the tool supports at least one among Amazon Lex, Dialogflow, and Rasa.
 \end{tablenotes}
\label{tab:all-tools}
\end{table}

\begin{itemize}[leftmargin=*]
\item \botium~\cite{botium}, which is a multi-platform state-of-the-practice test generation tool for task-based chatbots. It generates tests by \emph{statically} analyzing the implementation of the chatbots and deriving tests that use the chatbot training sentences to cover the possible dialog flows;
\item \charm~\cite{bravo2020testing}, which is a test generator tool that \textit{extends} \botium with the capability of modifying the sentences present in the tests, to exercise the robustness of the chatbot against alternative ways of asking the same questions present in the \botium tests;
\item \ctg~\cite{rapisarda2025test}, which is a \emph{dynamic} test case generation tool that exploits the tests initially generated by \botium to systematically explore the conversational space and embed oracles that reflect actual responses in the tests;
\item \asymobtcg (Test Case Generator)~\cite{canizares2024coverage}, which is a test generator tool integrated in the ASYMOB platform~\cite{lopez2022asymob,canizares2024measuring} that generates \botium tests according to \textit{dialog coverage metrics};
\item \tracer~\cite{del2025automated}, which is an automated LLM-based technique that synthesizes test profiles from systematic exploration of chatbot functionalities and uses them to guide user simulation.
\end{itemize}

\begin{table}[h!]
\renewcommand{\arraystretch}{1.2}
\setlength{\tabcolsep}{6pt}
\centering
\caption{Tools vs Frameworks.}
\label{tab:tools}
\begin{tabular}{l  c  c  c }
\toprule
& \textbf{Rasa} & \textbf{Dialogflow} & \textbf{Lex} \\
\midrule

\botium & \cmark & \cmark & \cmark \\
\hline

\charm & - & \cmark & - \\
\hline

\ctg & - & \cmark & - \\
\hline

\asymobtcg & \cmark & \cmark & - \\
\hline

\tracer & \cmark & - & -\\

\bottomrule
\end{tabular}
\end{table}

Table~\ref{tab:tools} summarizes the applicability of the existing test generation tools to the various platforms. Dialogflow chatbots will represent the common ground that can be used to compare and study the capabilities of most of the tools. On the other hand, \botium will be the primary source of evidence to compare how the same test generation strategies can adapt across different chatbot frameworks. 

\subsection{Methodology}
This section describes the methodology and the data collected to answer each research question.

\subsubsection{General Procedure}
To assess the capabilities of the test generation techniques, we plan to run each tool on each chatbot in our datasets, consistently with the compatibility constraints listed in Table~\ref{tab:tools}. The tools have different runtime, which depends on the strategy used to generate tests. Limiting the tools to the same runtime would cause the generation of incomplete test suites, since each technique explores the conversational space in a different order, and the results would be incomparable (e.g., a technique covering a part of the conversation totally ignored by the others, and vice-versa, only because the tool has been stopped early). We will thus run the techniques naturally, fully exploiting their capabilities, recording the time information, so that results can be discussed also in terms of the consumed resources.

Configuration-wise, we will set up the tools to generate the most complete tests. In particular, we will run the tools as follows: \botium using the \textit{multistepconvo} option, which enables the generation of multi-step conversations; \ctg with cleanup and teardown functions enabled to restore the original chatbot state after test execution~\cite{rapisarda2025test}; \charm with all input mutations enabled~\cite{bravo2020testing}; \asymobtcg with the \textit{exhaustive} objective that produces the tests with the highest coverage~\cite{canizares2024coverage}; and \tracer with maximum number of steps per session and \textit{nested-forward} option to produce a larger set of conversations. 

To guarantee the independence between test executions, we will set up and tear down the environment between test executions. 
During test executions, we will record the test outcomes (pass/fail), the executed conversations, and the logs produced by the tested chatbots and the test execution tools. Further, we will configure task-based platforms to log coverage data (e.g., to log the exercised intents). These artifacts will be used for the quantitative and qualitative analyses specific to each RQ, as described below.

\subsubsection{RQ1 (Test Correctness)}

\newtext{This research question \emph{measures} the relative portion of generated \emph{test cases} that are \emph{not executable} or \emph{not understandable} by the chatbot under test. A non-executable test is a test that includes syntactic errors that prevent it from being executed. A non-understandable test is a test that is syntactically correct but includes semantic errors that prevent it from being correctly processed by the chatbot, thus triggering a fallback intent that deviates from the intended test objective.}
For example, \botium may fail to capture dependencies between intents, generating tests that do not establish any conversation with the chatbot (e.g., skipping a mandatory salutation in the test).

To answer this research question, we will search for errors in the logs produced during test execution to compute the percentage of tests correctly generated. We will manually inspect and classify the errors, to derive insights about the causes of incorrect tests.

\subsubsection{RQ2 (Conversations)}

The range of conversations supported by a chatbot is mainly defined by (i) the intents implemented by the chatbot, (ii) the entities that can be used as part of the conversation, and (iii) the conversational flows. 

To answer this research question, we will measure the percentages of actual intents, entities, and flows exercised by each testing technique across all chatbots.  


In addition to coverage, we will study the capability of a test suite to exercise conversations using a fault-based approach. That is, we will iteratively inject mutants in the conversational part of each chatbot using \textsc{Mutabot}~\cite{clerissi2025towards,ferdinando2024mutabot}, which is a state-of-the-art tool for mutation testing in task-based chatbots. 
In this context, mutants are chatbots that incorporate single defects affecting a conversational element (e.g., a chatbot having a parameter removed from an utterance). To evaluate defect detection, we will measure the \emph{mutation score}, that is, the fraction of mutants detected by a test suite.
We plan to use all the mutant operators available in the experiment.

We will finally manually inspect the chatbots, the chats, and the tests to determine the elements of the conversations that are hard to test, and derive \emph{qualitative evidence} that can help direct future research effort.

\subsubsection{RQ3 (Functions)}
The functionalities of chatbots can be implemented in many different ways, ranging from custom backends developed with arbitrary technologies and languages, to external services (e.g., Google Calendar) simply invoked by chatbots. It is thus impossible to collect code coverage metrics or use mutation testing tools to analyze how well the functions used by the chatbots are exercised. 


Thus, we plan to collect data on how thoroughly backend services are exercised using the behavioral space coverage criterion introduced by Gazzola et al.~\cite{gazzola2023exvivomicrotest}, which measures the number of distinct sequences of \textit{N} operations that are executed, also accounting for the diversity of the parameter values used within calls. In particular, each data type is abstracted into a finite domain (e.g., positive integers can be abstracted into three classes ``0'', ``1'', and ``$>$1''), and operations executed with different parameter values (according to their abstracted domain) are counted as distinct operations in the coverage objectives. That is, \textit{book(1, ``doubleRoom'')} and \textit{book(5, ``doubleRoom'')} count as two distinct operations that can be executed. We will use the default abstractions associated with the data type, unless the semantics of the values require a refinement (e.g., it does not make sense to try to cover negative values for a positive-only variable). We will report the non-default abstractions used in the paper.

Also in this case, we will \emph{qualitatively} analyze the results by inspecting the cases of coverage items not covered by any testing approach, or covered uniquely by a testing approach, 
to derive a lesson learned about the open challenges in testing the functions implemented by the chatbots.

\subsubsection{RQ4 (Oracle)}
A key challenge is the automatic synthesis of program oracles that can be used to reveal failures, beyond crash-like failures, which are trivial to detect. 

We will derive the following metrics to assess the effectiveness of the oracles generated by each approach: \emph{True Positives} (number of failed tests correctly recognized as failed), \emph{False Positives} (number of passed tests incorrectly recognized as failed), \emph{False Negatives} (number of failed tests incorrectly recognized as passed); and \emph{True Negatives} (number of passed tests correctly recognized as passed).

We will execute the tests on the unchanged version of the bot to obtain passing tests, and we will exploit mutants to obtain failing tests. In both cases, we will manually inspect the conversations executed by the tests, and the chatbot implementation if necessary, to make sure that the passing and failing tests are truly passing and failing. 

We will derive the usual measures that combine the above metrics, such as \emph{accuracy} and \emph{f1-score}.
We will also classify the misbehaving tests (i.e., the tests with oracles that misclassify executions) to identify \emph{the soundly and particularly weak  oracles}.  

\subsubsection{RQ5 (Flakiness)}
We will finally investigate the flakiness of the generated tests. Indeed, there are many possible sources of flakiness for conversational agents, related to the use of natural language, but also technical challenges related to the setup of the chatbots under test, its configuration, and the interacting services. 

Since chatbot development platforms exhibit different behaviors and sources of flakiness (e.g., we observed cases of timeout errors raised by Rasa only when interacting with \botium), to address this RQ, we will execute the tests five
times on the same version of the chatbot and \emph{measure the consistency of the results}.
Based on our previous experiments with Dialogflow~\cite{ferdinando2024mutabot,rapisarda2025test}, and to ensure scalability over a larger subject size, we identified five iterations as a compromise between the cost of the study and the strength of the evidence.
Inconsistencies will be manually inspected to determine their causes, and finally report the \emph{main sources of flakiness} that testers have to consider when implementing tests for task-based chatbots.

\smallskip

For all the RQs, qualitative analysis will be based on the \emph{content analysis} method with two coders (two of the authors of this paper). In case coders disagree with any classification item (e.g., whether a failed test is a true or false positive), the coders will resolve the conflict in a dedicated meeting. In case the two coders cannot solve the conflict themselves, a third coder (a third author of this paper) will be involved in the discussion to finally resolve the conflict. The Cohen's Kappa, measuring the inter-rater reliability of the two coders, will be reported in the paper.

\section{Threats to Validity}\label{sec:threats}
The study is affected by internal, construct, and external threats to validity.

The main threats to internal validity concern the selection of the test generation tools and the chatbots. To mitigate the threat to the selection of the test generation tools, we systematically searched and selected all the tools designed to address task-based chatbots as defined in Section~\ref{sec:background}. Concerning the selection of the chatbots, since it was impossible to consider every possible platform available to develop task-based chatbots, we targeted the most popular ones~\cite{abdellatif2021comparison,perez2021choosing,motger2022software}, including both open-source and commercial platforms in the study.

The main threats to construct validity concern the setup of the chatbots and the significance of the metrics collected. Regarding the setup of the chatbots, we plan to mitigate this threat by carefully testing the correct integration of chatbots with any backend and external services before running any experiment
, and fix any technical issues we may find, preserving the original behavior as much as possible. Or, if fixing is not possible, replace the broken candidates with comparable alternatives from the curated datasets or even from the larger datasets of all chatbots available on GitHub in order to reach the intended subject size.

Concerning the significance of the metrics, we plan to collect a combination of coverage and fault-based metrics for both conversations and functionalities. This is likely to provide a fairly comprehensive picture of the effectiveness of test generation tools. Moreover, for each research question, we plan to complement the quantitative evidence with qualitative evidence that can provide insights about open research challenges and their causes. 

The main threats to external validity concern the selection of the subject chatbots. To mitigate this threat, we plan to choose the chatbots for the study by targeting the most significant ones that implement various tasks and use services and backend features. Combined with the use of chatbots implemented with different platforms of different natures, we expect the results to generalize well across task-based chatbots.

\section{Related Work}\label{sec:related}
The ubiquity of chatbots in everyday activities, being integrated into real-world applications, has raised the need for quality assurance techniques targeting conversational-based software, which has proven to be particularly challenging~\cite{cabot2021testing,li2022review,lambiase2024motivations}.

There are two main streams of work in delivering conversational-based software: approaches that heavily use LLMs and approaches exploiting NLP pipelines.
Although LLM-based chatbots are emerging, they are still in their infancy stage, and the services they deliver, on the one hand, hardly meet the reliability requirements demanded by certain products and, on the other hand, lack the level of control necessary to support tasks such as booking and user assistance.

In contrast, task-based chatbots rely on mature frameworks and approaches extensively used in industry~\cite{benaddi2024systematic,adamopoulou2020chatbots}, yet pose open challenges on quality assurance aspects~\cite{cabot2021testing,lambiase2024motivations}. This study thus focuses on task-based chatbots.

Over the years, \botium~\cite{botium} has consolidated itself as a multi-platform framework for test generation and execution of task-based chatbots, including both open-source and commercial solutions. The framework has put the foundations for several subsequent works, which have leveraged it for test augmentation, such as \charm~\cite{bravo2020testing} and \ctg~\cite{rapisarda2025test}, and the implementation of conversational coverage criteria, such as \asymobtcg~\cite{canizares2024coverage}. \botium has also been employed as the target of seminal papers on mutation testing for chatbots~\cite{ferdinando2024mutabot,gomez2024mutation,clerissi2025towards}.

Other approaches have applied black-box testing through user simulation in task-oriented dialog systems~\cite{vasconcelos2017bottester,bozic2019chatbot}. A recent work introduced \tracer~\cite{del2025automated}, a tool that infers a chatbot conversational model and synthesizes it into testing profiles to guide user simulation through LLMs. 

Input perturbation has also been the subject of investigation in conversational systems to assess their robustness~\cite{ruane2018botest,guichard2019assessing,iwama2019automated}, speech recognition capabilities~\cite{guglielmi2024help}, and security~\cite{bozic2020interrogating}. 
Since predicting the correctness of chatbot responses can be a complex task due to conversational variability, model-checking~\cite{silva2023modeling} and metamorphic testing~\cite{bozic2019testing,bovzic2022ontology} have  been employed to detect deviations from expected behavior and design NLP-based oracles, respectively. 

In preliminary experiments~\cite{ferdinando2024mutabot,rapisarda2025test,masserini2025brasato,clerissi2025towards,masserini2026assessing}, we observed that \botium shows limitations in generating articulated tests capable of exercising the conversational space, as well as in automatically defining flexible oracles beyond simple intent detection or exact text matching, and so do \charm~\cite{bravo2020testing} and \ctg~\cite{rapisarda2025test} tools exploiting it. 
More generally, most of the presented approaches have been evaluated from a small-scale perspective, relying on a limited number of subjects, outdated chatbots, toy examples, or chatbots built with non-standard technologies~\cite{masserini2025brasato}. These issues highlight the need for a more comprehensive and systematic study to compare state-of-the-art tools across multiple dimensions using larger datasets. 

\section{Conclusions}\label{sec:conclusions}

Task-based chatbots are systems that provide users with the possibility of accessing advanced services through conversational interfaces. The different nature of both the type of interface and the kind of interaction poses new challenges to test case generation. In fact, test generation techniques have to face the hard challenge of both generating and interpreting conversations.

This registered paper proposes to systematically study the effectiveness of test generation techniques on a curated selection of representative task-based chatbots implemented with heterogeneous platforms. The study covers five research questions concerning the correctness of the generated tests (RQ1), the capability of thoroughly exercising conversations (RQ2), the capability of activating the functionalities behind the conversational interface (RQ3), the strength of the generated oracles (RQ4), and finally the flakiness of the generated tests (RQ5). Overall, the answers to these five questions provide a fairly complete picture of the effectiveness of test generation techniques and the robustness of chatbot implementations. 

\bibliographystyle{ACM-Reference-Format}
\bibliography{references}

\end{document}